\documentclass[modern]{aastex62}

\definecolor{DarkGreen}{rgb}{0.0,0.4,0.0}  % define a custom color

\usepackage[normalem]{ulem}
\usepackage{soul}
\usepackage{cancel}

\newcommand{\kms}[1]{#1~$\text{km}\,\text{s}^{-1}$}
\received{March 9, 2022}
\revised{August 19, 2022}
\accepted{August 26, 2022}
\submitjournal{ApJ}
\shorttitle{Sigmoid Formation Through Loop Slippage}
\shortauthors{Pan et al.}
\begin{document}
\title{Sigmoid Formation Through Slippage of A Single J-shaped Coronal Loop}
\correspondingauthor{Rui Liu}
\email{rliu@ustc.edu.cn}
\author[0000-0002-0318-8251]{Hanya Pan}
\affiliation{CAS Key Laboratory of Geospace Environment, Department of Geophysics and Planetary Sciences, University of Science and Technology of China, Hefei 230026, China}
\affiliation{Mengcheng National Geophysical Observatory, School of Earth and Space Sciences, University of Science and Technology of China, Hefei 230026, China}
\author[0000-0003-0510-3175]{Tingyu Gou}
\affiliation{CAS Key Laboratory of Geospace Environment, Department of Geophysics and Planetary Sciences, University of Science and Technology of China, Hefei 230026, China}
\affiliation{CAS Center for Excellence in Comparative Planetology, Hefei 230026, China}
\author[0000-0003-4618-4979]{Rui Liu}
\affiliation{CAS Key Laboratory of Geospace Environment, Department of Geophysics and Planetary Sciences, University of Science and Technology of China, Hefei 230026, China}
\affiliation{CAS Center for Excellence in Comparative Planetology, Hefei 230026, China}
\affiliation{Collaborative Innovation Center of Astronautical Science and Technology, Harbin, China}
\begin{abstract}
A well-known precursor of an imminent solar eruption is the appearance of a hot S-shaped loop, also known as \emph{sigmoid}, in an active region (AR). Classically, the formation of such an S-shaped loop is envisaged to be implemented by magnetic reconnection of two oppositely oriented J-shaped loops. However, the details of reconnection are elusive due to weak emission and subtle evolution during the pre-eruptive phase. In this paper, we investigate how a single J-shaped loop transforms into an S-shaped one through the slippage of one of its footpoints in NOAA AR 11719 on 2013 April 11. During an interval of about 16 min, the J-shaped loop slips through a low-corona region of strong electric current density in a bursty fashion, reaching a peak apparent speed as fast as over \kms{1000} at the slipping footpoint. The enhancement of electric current density, as suggested by non-linear force-free field modeling, indicates that the ``non-idealness'' of coronal plasma becomes locally important, which may facilitate magnetic reconnection. The loop segment undergoing slipping motions is heated; meanwhile, above the fixed footpoint coronal emission dims due to a combination effect of the lengthening and heating of the loop, the latter of which is manifested in the temporal variation of dimming slope and of emission measure. These features together support an asymmetric scenario of sigmoid formation through slipping reconnection of a single J-shaped loop, which differs from the standard tether-cutting scenario involving a double J-shaped loop system.
\end{abstract}
\section{Introduction} \label{sec:intro}
Composed of a group of coherently twisted magnetic field lines, magnetic flux ropes are an important structure in many models that have illuminated the key physics of solar eruptions \cite[see the recent reviews by][and references therein]{Liu2020,Patsourakos2020}.
However, the actual processes leading to the formation of magnetic flux ropes and the subsequent onset of eruptions remain elusive. 
A flux rope may form in the convection zone and then emerge through the photosphere into the corona \cite[]{Fan2003,Cheung2014}. The emerging process, however, can be severely hampered by heavy plasma trapped at the bottom of the rope. Alternatively, a flux rope may form in the corona via the so-called `tether cutting' reconnection \citep{Moore1980, Moore2001}. 
This process can be driven by photospheric shearing and converging flows around the polarity inversion line \cite[PIL;][]{Ballegooijen1989}; consequently, magnetic shear increases near the PIL, which is often manifested as the presence of a pair of opposing J-shaped loops. Through magnetic reconnection between their inner legs, the double J-shaped loops are converted into a longer twisted flux rope and a shorter loop that consequently submerges, exhibiting flux cancellation in the photosphere \cite[]{Ballegooijen1989,Green2009}. 
\par
The tether-cutting scenario was motivated by the evolution of an S-shaped feature known as `sigmoid', observed in soft X-rays (SXRs) or extreme-ultraviolet (EUV). Soon after being discovered by the Soft X-Ray Telescope on-board Yohkoh \citep{Sakurai1992}, sigmoids are recognized as an important CME progenitor \citep{Canfield1999}. Despite the low time cadence, Yohkoh images demonstrate the sigmoid-to-arcade transformation \citep{Sterling2000, Moore2001}, i.e., a double J-shaped loop bundle is observed before eruptions whereas a cusp-shaped flare loop or arcade in the wake of eruptions. With high-cadence, high-resolution SDO/AIA images, \citet{Liu2010} observed a complete transformation process from a double-J-shaped loop bundle to a continuous S-shaped loop that subsequently deforms ceaselessly to an arc-shaped eruptive front. With a nonlinear-force-free field (NLFFF) model, \citet{Liu2013} traced magnetic field lines from four brightening kernels in AIA 1700~{\AA} and found that the changes of magnetic-connectivity before and after the flare conform to the tether-cutting model. 
\par
It is natural to connect sigmoids with magnetic flux ropes or highly sheared magnetic arcades, since the central section of a sigmoid is typically aligned with the photospheric PIL, indicating enhanced magnetic stresses and electric currents therein.
\citet{Rust1996} measured the aspect ratio (width to length) of 103 transient, bright sigmoids in X-ray images taken by Yohkoh. They found that the peak distributed aspect ratio in observation is consistent with the one obtained from a kinked flux rope model. \citet{Mckenzie2008} reported a sigmoid lasting for about 3 days, comprising two separate J-shaped loop bundles. The persistence of the sigmoid and the double-J configuration provide strong support for the bald-patch separatrix surface model, and hence for the existence of a kinked flux rope as argued by \citet{Green2009}. Sigmoids have since been considered as a proxy of magnetic flux ropes. 
\par
Built upon the two-dimensional (or 2.5-dimensional) `standard' flare model \citep{Carmichael1964,Sturrock1966,Hirayama1974,Kopp&Pneuman1976}, the tether-cutting model still depicts a classical ``cut-and-repaste'' scenario. But in reality, magnetic reconnection exhibits complex three-dimensional (3D) structures \cite[]{Priest1995,Priest2003,Pontin2011}. 3D reconnections take place not just at topological separatrices and null points but in quasi-separatrix layers \cite[QSL;][]{Demoulin2006}, where magnetic field lines continually exchange magnetic connectivities with neighboring lines, exhibiting flipping or slipping motions of their footpoints. \cite{Janvier2013} demonstrated through MHD simulations that the temporal variation of slippage speed is strongly correlated with the spatial structure of magnetic connectivity gradient, the latter of which can be quantified by the mapping norm $N$ \cite[]{Priest1995} or squashing factor $Q$ \cite[]{Titov2002}. Typically $N$ and $Q$ increase toward the center of QSLs, approaching infinity at where separatrices or nulls are embedded \cite[e.g.,][]{Liu2016,Chen2020}. Many observational studies report slipping motions in SXRs and EUV \cite[e.g.,][]{Aulanier2007, Li2014, Dudik2014, Li2015,Dudik2016,Gou2016}, with speeds ranging from several tens to slightly over one hundred kilometers per second. The temporal variation of slippage speed, however, is difficult to obtain in observation, despite that it may provide a glimpse of magnetic structures in reconnection regions.\par
In this paper, we investigate an EUV sigmoidal loop formed shortly before the flare onset. It evolves from a single J-shaped loop through footpoint slippage, rather than from a double J-shaped loop system. The difference from the tether-cutting scenario may shed new light on the formation mechanism of magnetic flux ropes. In the remainder of the paper, we present the observations and data analysis in Section \ref{sec:obs},  and discuss these results in Section \ref{sec:disc}.
\section{Observation and Analysis} \label{sec:obs}
\subsection{Instruments and Methods}\label{instru}
In this study we use EUV images taken by the Atmospheric Imaging Assembly  \citep[AIA;][]{Lemen2012} on board the Solar Dynamics Observatory \citep[SDO;][]{Pesnell2012} at a spatial scale of $0''.6$ pixel$^{-1}$ and a cadence of 12~s. We focus on two hot passbands (131~{\AA} and 94~{\AA}), in which the slippage of the coronal loop is visible. We also use other AIA passbands, 335, 211, 193, and 171~{\AA}, to study coronal dimming, and the AIA 304~{\AA} passband along with H$\alpha$ images taken by the Kanzelh\"{o}he Solar Observatory (KSO) and by the Global Oscillation Network Group (GONG) to examine the filament in the same AR. To diagnose the plasma temperature in the optically thin corona, we employ an algorithm based on sparsity \cite[]{Cheung2015,Su2018} to perform differential emission measure (DEM) inversions on coronal images obtained by AIA's six EUV passbands.
\par
Photospheric magnetic field is observed by the Helioseismic and Magnetic Imager \mbox{\citep[HMI;][]{Hoeksema2014}} on board SDO. We use HMI vector magnetograms (data product hmi.sharp\_cea\_720s) to extrapolate non-linear force-free field (NLFFF) with an optimization approach \cite[]{Wiegelmann2004,Wiegelmann2012} and to derive flow field by the Differential Affine Velocity Estimator for Vector Magnetograms \citep[DAVE4VM;][]{Schuck2008}.
\subsection{Overview} \label{subsec:overview}
\begin{figure*}[ht!]
	\centering\includegraphics[width=\textwidth]{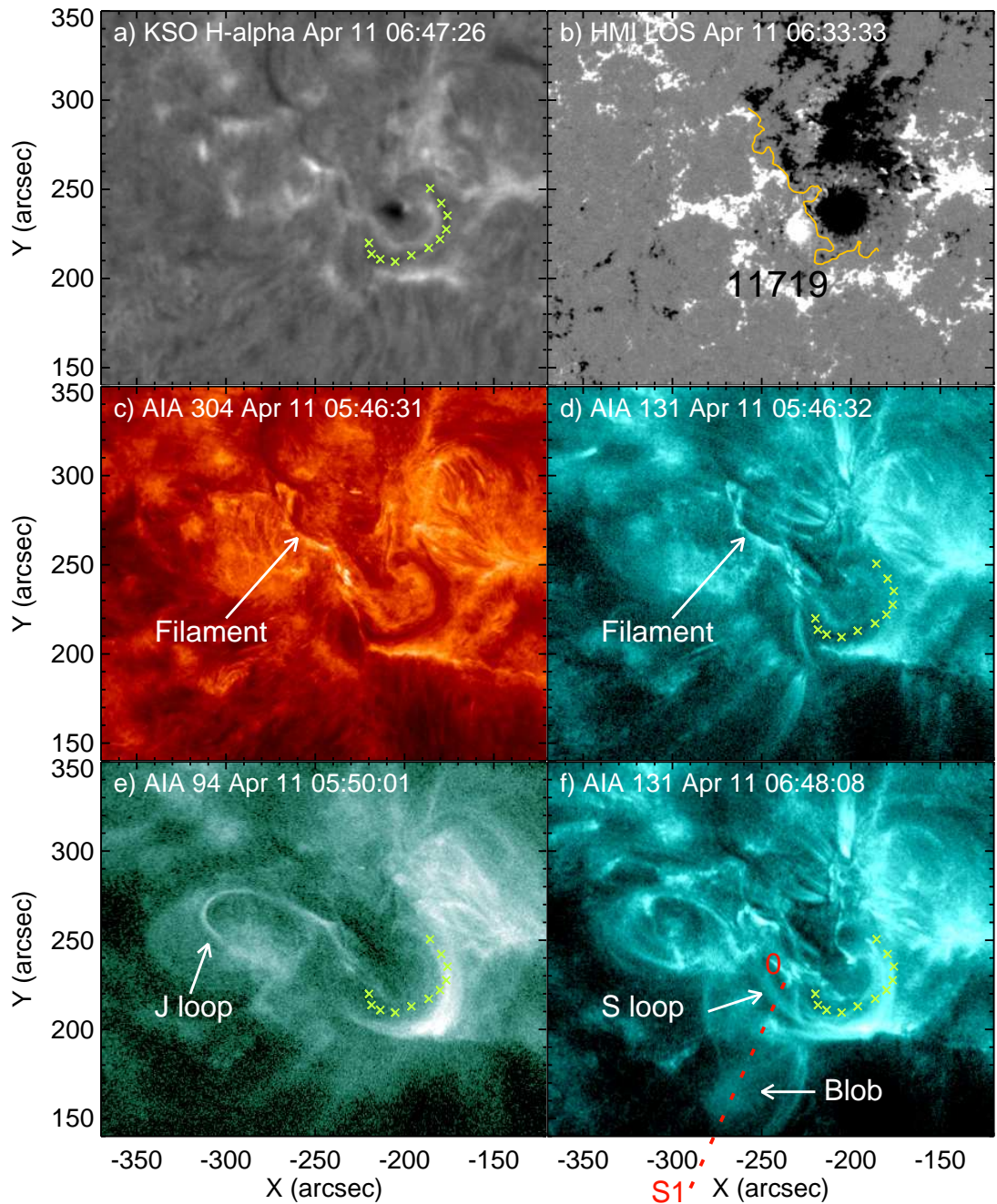}
	\caption{Overview of the eruption on 2013 April 11. 
		Panel (a) shows the filament in H$\alpha$ taken by KSO. Panel (b) shows the HMI line-of-sight magnetogram of the same region, superimposed by the PIL segment (orange curve) relevant to the eruption. Panels (c)-(d) show the filament in 304~{\AA} and 131~{\AA} respectively. Panels (e)-(f) show the transformation from a J-shaped to an S-shaped loop. The `x' symbols mark the position of a semicircular filament segment extracted from H$\alpha$ in panel (a). A virtual slit S1, through which we made the stack plot in Figure~\ref{fig:tl}(d), is marked in panel (f) with its starting position indicated by `0'. An animation of AIA 131, 94, and 304~{\AA} original and running difference images is available online, covering the time interval of 05:30--08:00 UT. The animation includes a panel at the top to show the GOES 1--8~{\AA} light curve, with an animated vertical line indicating when the images were taken.
		\label{fig:ovv}}
\end{figure*}

The eruption of interest is a GOES-class M6.5 two-ribbon flare in NOAA AR 11719 (Figure \ref{fig:ovv}(b)) on 2013 April 11, associated with a halo CME directed at the Earth. The flare started at 06:55 UT (Figure \ref{fig:tl}(b)) according to GOES 1--8~{\AA} flux. AR 11719 was featured by a large sunspot of negative polarity in the center, surrounded by small spots and pores of mixed polarities (Figure \ref{fig:ovv}(b)). To the south and east of the large sunspot was a filament aligned along the PIL, with its southern branch half circling the sunspot and the northern branch presenting a sinuous shape (Figure \ref{fig:ovv}(a \& c)). Closely following the semicircular filament branch was a bright loop bundle, clearly visible in relatively hot passbands, 131, 94, and 335~{\AA} (e.g., Figure \ref{fig:ovv}(e)). In H$\alpha$ images, the filament was long-lasting and remained stationary throughout the eruption (Figure~\ref{fig:fl}); however, the filament material was observed to flow along the loop when it was detached from the filament (\S\ref{subsubsec:counterstreaming}). The semicircular loop bundle was also stable prior to the eruption, but was obscured by post-flare loops during the eruption (see the animation accompanying Figure~\ref{fig:ovv}). \par
At 05:46 UT, over one hour before the flare onset, the sinuous filament segment brightened in EUV (e.g., Figure \ref{fig:ovv}(c \& d)), while the rest of the filament remained dark. The brightening lasted for about 5 min. This observation indicates that the sinuous segment was relatively independent of the rest of the filament. Apparently crossing the sinuous filament segment from above, a J-shaped coronal loop became visible at about 05:50 UT in 94~{\AA} (Figure \ref{fig:ovv}(e)) and 131~{\AA} (not shown). From 06:28 to 06:46 UT, the western footpoint of the J-shaped loop, originally located next to the eastern end of the semicircular filament (marked by `x' symbols) slipped through a curved trace around the major sunspot till near the western end of the semicircular filament. Consequently, the single J-shaped loop was transformed into an S-shaped structure (also referred to as sigmoid; Figure \ref{fig:ovv}(f)) through the slipping process. Starting from about 05:55 UT, a bundle of cusp-shaped loops rose slowly and rotated into a helical shape, whose top section morphed into a diffuse blob at about 06:25 UT, 3 minutes before the onset of the J-shaped loop's slippage. The blob of hot material then expanded and eventually erupted with the sigmoid (Figure \ref{fig:ovv}(f) and accompanying animation).  \citet{Vemareddy2014} studied the blob's evolution in detail and interpreted it as a kinked flux rope.
\subsection{Loop Slippage and Slipping Speed} \label{subsubsec:slip_fp}
%TS
At 06:26 UT,  the J-shaped loop began to brighten in hot passbands, 131 and 94~{\AA}. Two minutes later, the western footpoint of the loop started to slip westward along the PIL (Figure \ref{fig:slip}). The time range of the slipping process is marked with two red dashed lines in Figure \ref{fig:tl}(b)-(d). Below, we took a series of steps to investigate the evolution of the loop, especially of its slipping footpoint. \par
First, to enhance the visibility of the loop undergoing slipping motions, we paired 131~{\AA} and 94~{\AA} images that are closest in time stamp and added them in a ratio that is optimized through trial and error. The running difference images of 131~{\AA} and 94~{\AA} passbands are also combined in a similar fashion to highlight the slipping motion. Then, we marked the loop in running difference images using `x' symbols (Figure \ref{fig:slip}), with the aid of the original images to locate the eastern section of the loop that was relatively diffuse. During the early phase of the slippage, the western section of the loop appeared to be attached to the sinuous filament segment (Figure \ref{fig:slip}(a--f)). But later, the loop was `torn' off from the filament while the loop's western footpoint accelerated to slip away (Figure \ref{fig:slip}(g--l)). By contrast, the eastern section of the loop was kept relatively fixed. Finally, we calculated the slipping speed from the western footpoint's displacement observed in 14 time instants (only six of them are plotted in Figure \ref{fig:slip}). The initial slipping speed was about 20 km~s$^{-1}$, then increased to over 1000 km~s$^{-1}$, and finally slowed down when the slipping footpoint approached the sunspot. The peak speed is at least one order of magnitude faster than those reported in previous studies \citep[e.g.,][]{Aulanier2007,Dudik2014,Li2014,Li2015,Dudik2016}. 
\par
To estimate the uncertainty in our pinpointing the slipping footpoint, we checked the intensity variation across the footpoint. An example is shown in the inset of Figure~\ref{fig:slip}(j). We first located the local maximum in the footpoint region. In practice, this is done by enclosing the footpoint with a relatively large box (not shown) and then searching for the pixel with maximum DN within the box. Centering around this pixel, we obtained the averaged intensity over 5 pixels across the footpoint region in the $X$- (light green) and $Y$-direction (dark green), respectively. The full width at half maximum in the $X$- and $Y$-direction are shown by blue rectangles in Figure 4, which indicate the double of measurement uncertainties adopted to derive the error bars for slipping speeds (Figure \ref{fig:tl}(e)), following the propagation of uncertainties. In some cases one can only determine half width at half maximum at one side, which is then taken as the measurement uncertainty.
\par
To understand the loop slippage in relation to the magnetic configuration in AR 11719, we calculated the current density in the reconstructed NLFFF by $\mathbf{J}=\frac{1}{\mu_0}\nabla\times\mathbf{B}$ under the magnetohydrodynamics assumption. Further, we integrated the absolute value of current density within the height range of 2.9--47.4 Mm (or 4--65 pixel above the surface in our NLFFF), i.e., from the top of the chromosphere extending upward until about one hydrostatic scale height of the 1-MK corona \cite[][Eq.(3.1.16)]{Aschwanden2006}, to obtain the 2-D distribution of ``column'' current density in units of $\text{A}\,\text{m}^{-1}$ in the low corona (Figure \ref{fig:mag_cur}(b)). We then superimposed the extracted positions of the slipping loop, which are color-coded by observed time, over the map of column current densities. One can see that the trace of the slipping footpoint as marked by asterisks mainly ran across the region with intense coronal current densities. As a sanity check, we found that in the selected height range current densities in the region of interest significantly exceed those in the background (represented by a reference region of weak current densities; Figure~\ref{fig:mag_cur}(c)). It is remarkable that in projection photospheric regions with strong vertical current densities $J_z$ (Figure~\ref{fig:mag_cur}(a)) can be significantly displaced from coronal regions with strong column current densities  (Figure~\ref{fig:mag_cur}(b)).
\subsection{Asymmetric Heating}
As a result of the slipping motion, a sigmoidal loop was produced (marked by plus symbols in Figure \ref{fig:avet}) with temperatures as high as over $\log_{10} T\mathrm{[K]}=6.7$. The EM-weighted mean temperature (Figure \ref{fig:avet}(b)) was higher in the western part of the sigmoidal loop than in the eastern part. One caveat is that the DEM analysis could be compromised by the optically thick filament material (black solid curve in Figure \ref{fig:avet}(b)) that overlapped the sigmoidal loop along the line of sight, because the filament material appeared in absorption in cold passbands but was obscured by brightening structures in hot passbands. However, we noticed that the temperature enhancement is also present along the loop segment with no filament material obstructing the line of sight. 
\par
\begin{figure*}[ht!]	
	\centering
	\includegraphics[width=0.9\textwidth]{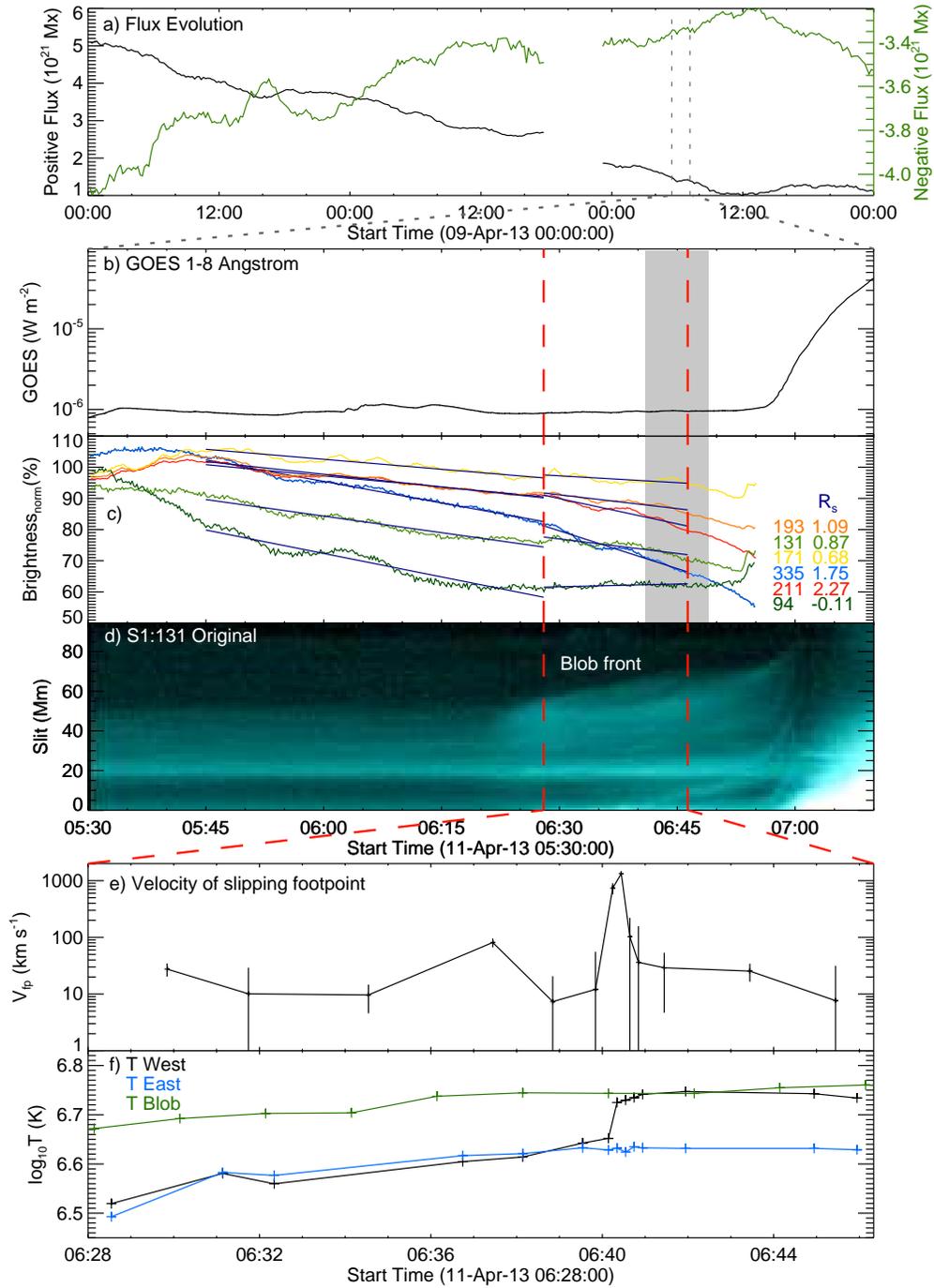}
	\caption{Temporal evolution during the pre-flare phase. 
		Panel (a) shows the evolution of magnetic flux from the rectangular region marked in Figure~\ref{fig:fv}(a) over 3 days prior to the flare of interest. Panels (b)-(d) zoom in on a 2-hr interval before and at the flare onset to show GOES 1-8~{\AA} flux, normalized brightness of six AIA EUV passbands averaged over the elliptical region marked in Figure \ref{fig:dim}, and the evolution of coronal structures seen through the slit S1 marked in Figure \ref{fig:ovv}(f). In panel (c), the ratio $R_S$ of dimming slope during the slipping phase over that during the pre-slipping phase is annotated for each passband (\S\ref{subsubsec:asym_dim}). The shaded bar indicates the time range during which counter-streaming flows were observed (\S\ref{subsubsec:counterstreaming}).
		Panel (e) shows the evolution of the slipping speed with estimated error bars (\S\ref{subsubsec:slip_fp}).
		Panel (f) shows the average temperature of the eastern and western segments of the S-shaped loop and of the blob. The segments are indicated by yellow curves in Figure~\ref{fig:avet}(a). 
	   \label{fig:tl}
		}
\end{figure*}
\begin{figure*}[ht!]
	\plotone{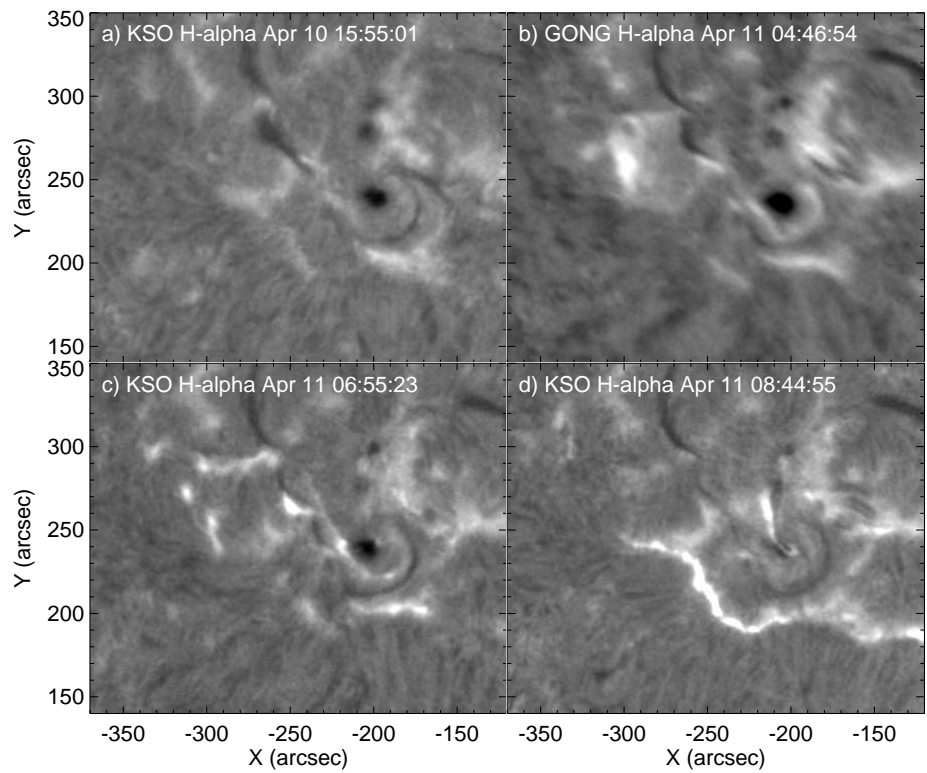}
	\caption{Filament in different periods.
		Panels (a) \& (b) were observed before the loop slippage. 
		Panel (c) was observed after the loop slippage but right at the flare onset.
		Panel (d) was observed during the flare gradual phase.
		\label{fig:fl}}
\end{figure*}
\begin{figure*}[ht!]
	\centering\includegraphics[width=\textwidth]{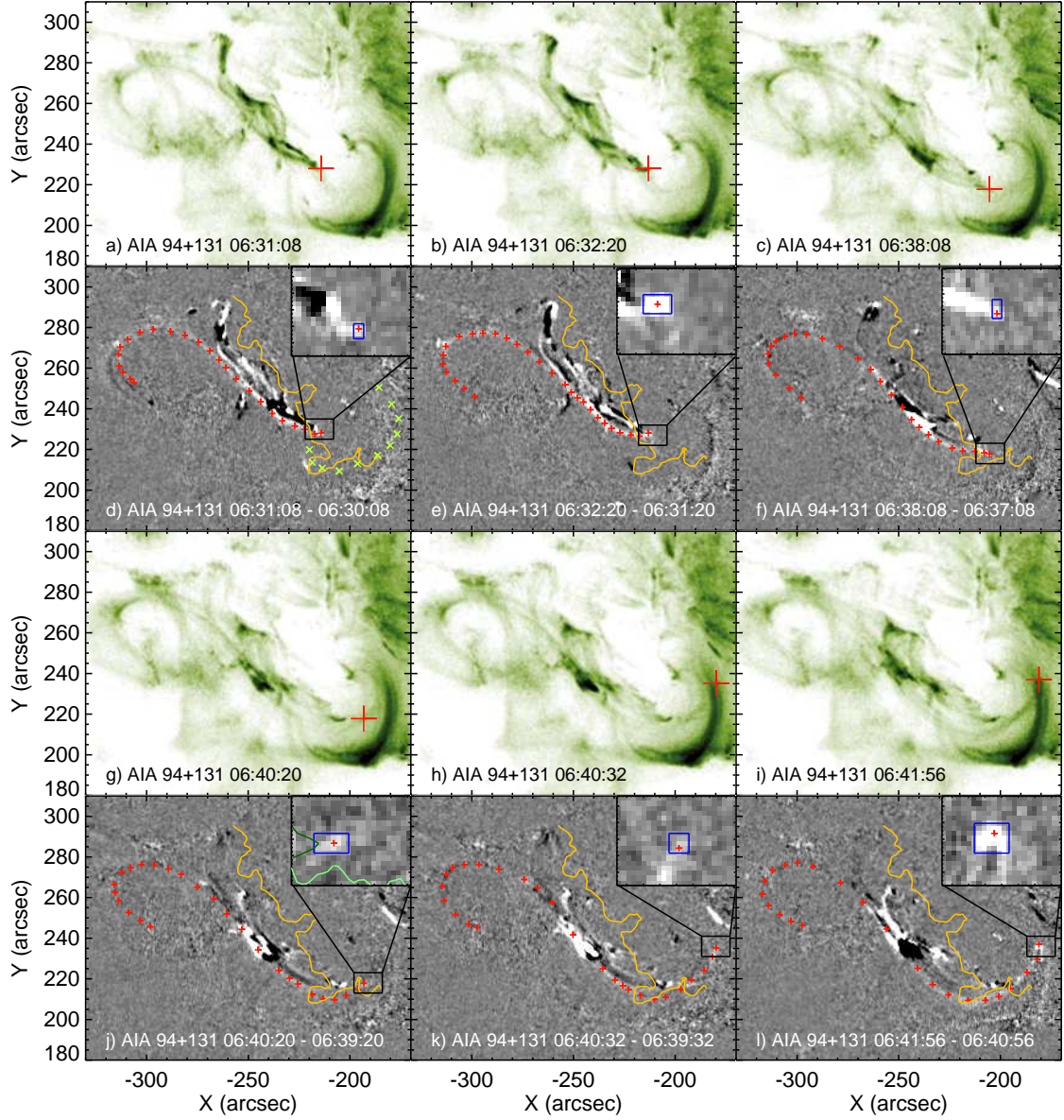}
	\caption{Transformation from the J-shaped to S-shaped loop.
		Panels (a)-(c) \& (g)-(i) show composite 131~{\AA} and 94~{\AA} images (see \S\ref{subsubsec:slip_fp}), with the plus symbol indicating the western footpoint of the loop. Panels (d--f) \& (j--l) show corresponding running difference images, superimposed by the PIL (orange) derived from the HMI line-of-sight magnetograms, same as in Figure~\ref{fig:ovv}b. The loop under slipping motions is marked by red plus symbols. The green `x' symbols in (d) indicate the semicircular filament segment, same as in Figure~\ref{fig:ovv}. The time stamp in each panel is given by the 131~{\AA} passband. The insets in panels (d--f) \& (j--l) zoom into the rectangular region around the loop's western footpoint. In panel (j) the light and dark green curves show the average intensity variation across the footpoint region in the $X$- and $Y$-direction, whose full width at half maximum is taken as the double of measurement uncertainty $\sigma_X$ and $\sigma_Y$ of the footpoint position, respectively (see \S\ref{subsubsec:slip_fp}). The uncertainties are shown as the blue rectangle in each inset.  An animation of the composite 131~{\AA} and 94~{\AA} images is available online, covering the time interval of 06:28--07:10 UT. The animation includes a panel at the top to show the GOES 1--8~{\AA} light curve, with an animated vertical line indicating when the images were taken.
		\label{fig:slip}}
\end{figure*}
\begin{figure*}[ht!]
	\centering\includegraphics[width=\textwidth]{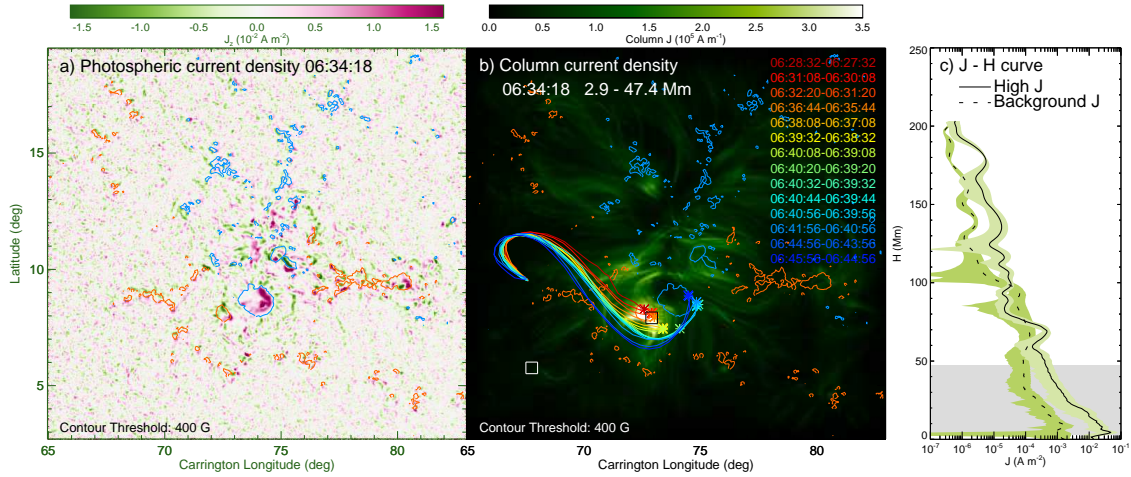}
	\caption{Current density distribution in the corona in relation to the loop slippage. Panel (a) shows the vertical component $J_z$ of current densities derived from HMI vector magnetograms (hmi.sharp\_cea\_720s data products) during the loop slippage. Vector magnetograms are smoothed by a $4\times4$ pixel$^2$ running box before $J_z$ is calculated. Panel (b) shows the integration of the absolute value of column current density over the height range from 2.9 to 47.4 Mm. The orange and blue contours in (a) and (b) indicate $B_z$ component of photospheric magnetic fields at $\pm400$ G, respectively. The color-coded curves show the loop position obtained from the composite of 131 and 94~{\AA} running difference images at different time instants. Panel (c) gives the variation of electric current density with increasing height in a representative region of strong current density (black box in (b)) and in a reference region of weak current density (white box in (b)). The green shaded range indicate the standard deviation of current densities in the selected box regions. The gray shaded bar indicates the height range within which the column current density in (b) is obtained.
		\label{fig:mag_cur}}
\end{figure*}
\begin{figure*}[ht!]
	\plotone{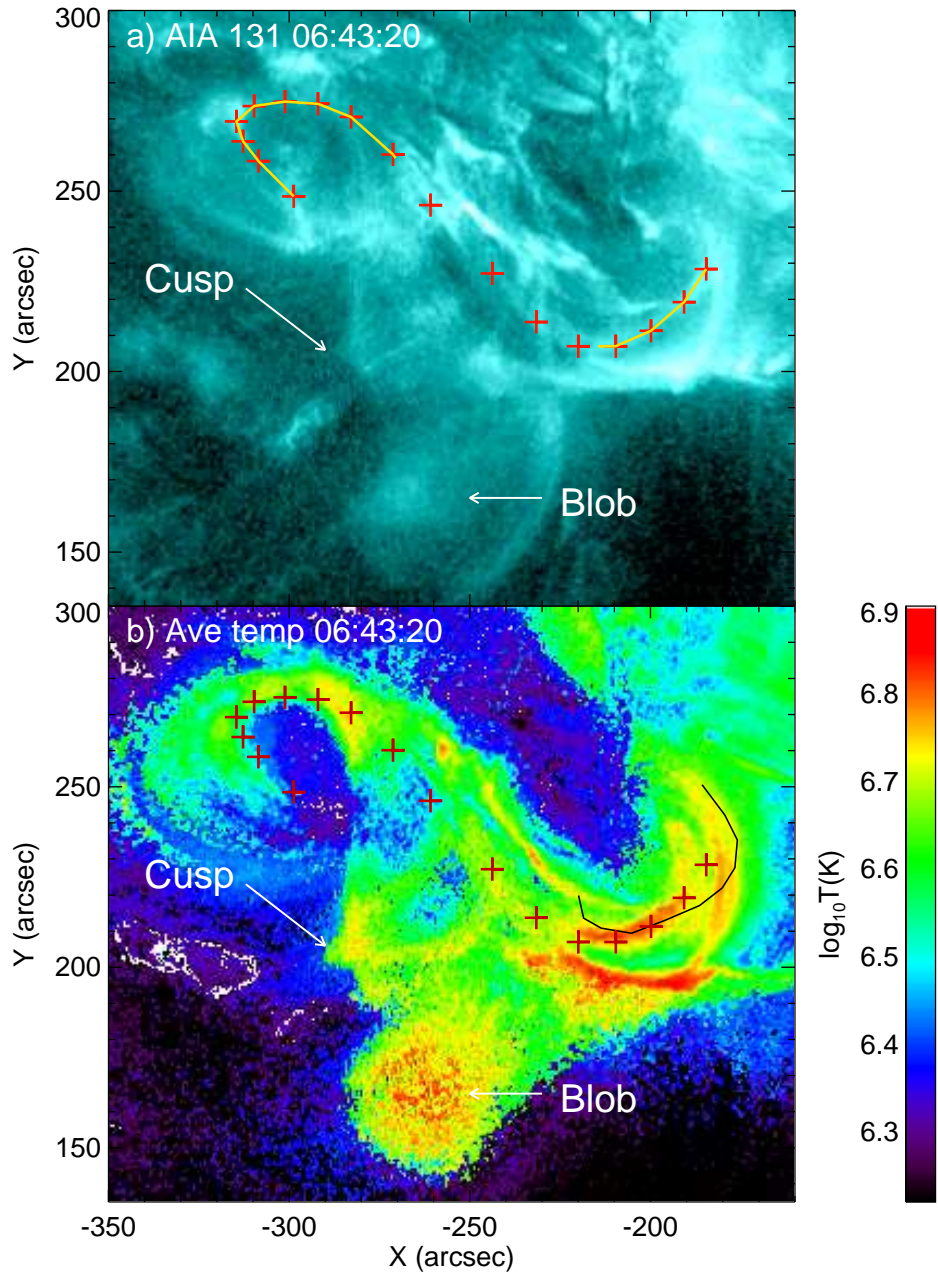}
	\caption{Temperature distribution. 
		(a) AIA 131~{\AA} image taken at 06:43 UT when the sigmoidal loop (marked by red plus symbols) was fully fledged through slipping reconnection. Two yellow curves indicate where we sampled the temperature of the eastern and western segments, respectively, of the sigmoidal loop (Figure~\ref{fig:tl}(f)). (b) Map of the DEM-weighted mean temperature.  The black curve indicates the position of the semicircular filament segment extracted from Figure \ref{fig:ovv}(a). An animation of 131~{\AA} images and temperature maps is available online, covering the time interval of 05:30--07:00 UT. The animation includes a panel at the top to show the GOES 1--8~{\AA} light curve, with an animated vertical line indicating when the images were taken. %\comment{use different symbols for the filament in (b)}
		\label{fig:avet}}
\end{figure*}

Further, we take the average of temperatures sampled at the extracted positions at different time instants (Figure \ref{fig:mag_cur}(b)) from a eastern and western segment of the sigmoidal loop, respectively; each segment roughly covers the curved end of the S-shaped loop (yellow curves in Figure \ref{fig:avet}(a)). The average temperatures at both segments were roughly equivalent and both increased gradually during the early slipping phase, but at the western segment the temperature suddenly jumped from $\log T=6.6$ to 6.7 at 06:40 UT (Figure~\ref{fig:tl}(f)), coinciding with the peak time of the slipping speed (Figure~\ref{fig:tl}(e)). Hence we argued that the heating at the western, slipping section of the sigmoidal loop is not a spurious effect, but associated with the slipping process.
\par
In comparison, we sampled the temperature of the blob from an $8''\times8''$ rectangle in the center of the blob at 10 time instants. One can see that the blob was heated gradually during the slipping process (Figure~\ref{fig:tl}(f)), and had a higher temperature than the sigmoidal loop until the slipping speed peaked at about 06:40 UT, when the western section of the sigmoidal loop was heated to a similar temperature. The different temperature-evolution profiles suggest that the blob and the sigmoidal loop were most likely independent magnetic structures. 
\par
This inference can be further corroborated by investigating the evolution of the blob. In the original and running difference images of AIA 131 and 94~{\AA} passbands (see the animation accompanying Figure~\ref{fig:ovv}), one can see a cusp-shaped loop bundle appeared as early as 05:35 UT, almost one hour before the loops slippage, and later split into two major bundles at about 06:00 UT, due to the slow rising and expanding of the western bundle. It further rotated into a helical shape whose top section expanded into the diffuse blob. The blob was anchored similarly as the cusp-shaped loop, with one footpoint near the middle of the sinuous filament segment and the other in the neighborhood of the western end of the semicircular filament segment.  
\subsection{Asymmetric Dimming} \label{subsubsec:asym_dim}
\begin{figure*}[ht!]
	\centering\includegraphics[width=\textwidth]{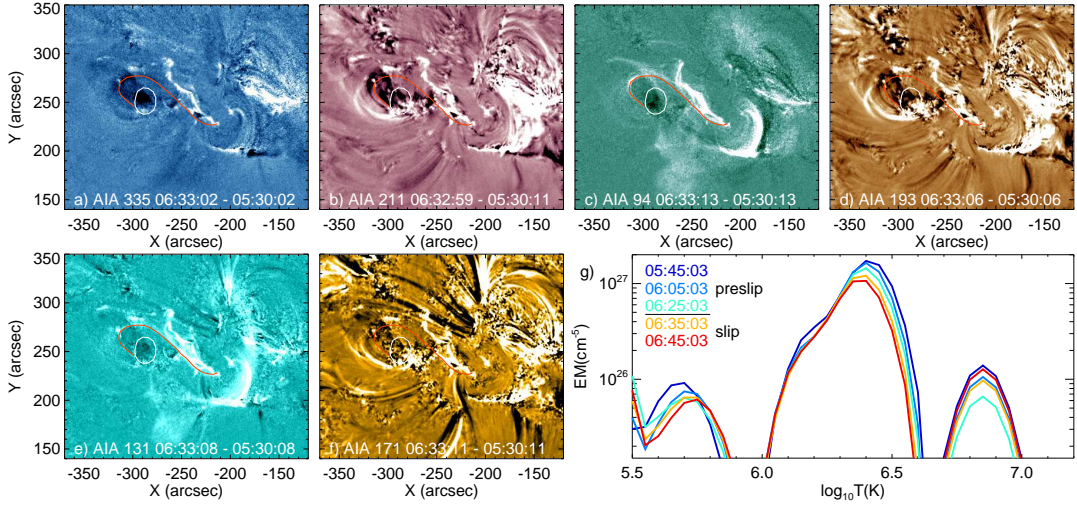}
	\caption{Asymmetric dimming. 
		Panels (a)-(f) show base difference images of six AIA EUV passbands. The normalized brightness of the six EUV passbands from the elliptical area in the dimming region is plotted in Figure \ref{fig:tl}(c). The red curve indicates the sigmoidal loop, extracted from the image taken at 06:32 (Figure \ref{fig:slip}(e)). Panel (g) gives the $\mathrm{EM}=\mathrm{DEM}\cdot\Delta T$ averaged over the elliptical dimming region at different time instants (color coded) within the temperature range $\log_{10}T[\text{K}]=[5.5,\,7.1]$, which is divided into logarithmically uniform intervals of $\log_{10}\Delta T=0.05$ in the DEM analysis.		
		\label{fig:dim}
	}
\end{figure*}

About 1 hour before the flare onset, the area around the eastern footpoint of the J-shaped loop system started to dim in both hot and cold passbands (Figure \ref{fig:dim}). We acquired the evolution of dimming by taking the average intensity in one quasi-elliptical area around the eastern footpoint. The obtained curves are plotted in Figure \ref{fig:tl}(c) in different colors. One can see that the radiation from this area kept decreasing during the pre-flare phase, which is termed `pre-eruption dimming'. This phenomenon has been reported in a few studies \citep{Gopalswamy1999,Zhang2017,Qiu2017,Wang2019,Pan2021}, in which dimming was detected as early as 1 to 5 hours before a flare/CME onset. It is generally explained as density depletion due to the gradual expansion of the eruptive structure during its slow rise. 
\par
Generally, the dimming depth and slope differ among different passbands. In our case, we noticed that in 94, 335, and 211~{\AA} the dimming slope changes greatly from before to during the loop slippage. To quantify this, we obtained the average dimming slope $\overline{S}$ from the slope of linear fitting to the intensity lightcurves over the time interval before ($\overline{S}_\mathrm{preslip}$) and during the loop slipping phase ($\overline{S}_\mathrm{slip}$). The ratio  $R_S=\overline{S}_\mathrm{slip}/\overline{S}_\mathrm{preslip}$ is given in Figure~\ref{fig:tl}(c). The $R_S$ values are consistent with our impression that during the slipping phase, the dimming slope is significantly enhanced in 335 and 211~{\AA}, but depressed in 94~{\AA}, compared with the pre-slipping phase. $R_S$ even turns negative in 94~{\AA} because of a slight recovery in intensity during the slipping phase. 
\par
With the DEM analysis, we found that the EM of the dimming region exhibits a complex temporal variation in its `hot bump' at about $\log_{10}T[\mathrm{K}]\sim 6.85$, which is the formation temperature of the \ion{Fe}{18} emission line ($\lambda=93.93\,$\AA) that dominates the 94~{\AA} passband \cite[Figure~\ref{fig:dim};][]{ODwyer2010}. The variation indicates that the amount of hot mateiral at $\sim\,$6~MK decreases during the pre-slipping phase but increases during the slipping phase. On the other hand, the temporal variation of the `warm bump' at about $\log_{10}T[\mathrm{\mathrm{K}}]\sim 6.4$, which includes both \ion{Fe}{16} ($\log_{10}T[\mathrm{K}]=6.45$, $\lambda=335.41\,$\AA) and \ion{Fe}{14} ($\log_{10}T[\mathrm{K}]=6.3$, $\lambda=211.32\,$\AA) that dominates the 335 and 211~{\AA} passband, respectively, indicates that the amount of warm material at $\sim\,$2-3~MK keeps decreasing during the whole time interval of dimming. These are consistent with the temporal variation of dimming slope among different passbands, especially for 94, 335, and 211~{\AA} (Figure~\ref{fig:tl}(c)).  
\par
Thus, we speculate that the dimming at the eastern footpoint of the sigmoidal loop might be caused by the loop expansion during the pre-slipping phase, but by a combination of the loop lengthening and heating during the slipping phase. In contrast to the eastern footpoint, no obvious dimming was seen around the western footpoint during the whole time before the flare. 
\subsection{Counter-Streaming Flow} \label{subsubsec:counterstreaming}
\begin{figure*}[ht!]
	\centering\includegraphics[width=\textwidth]{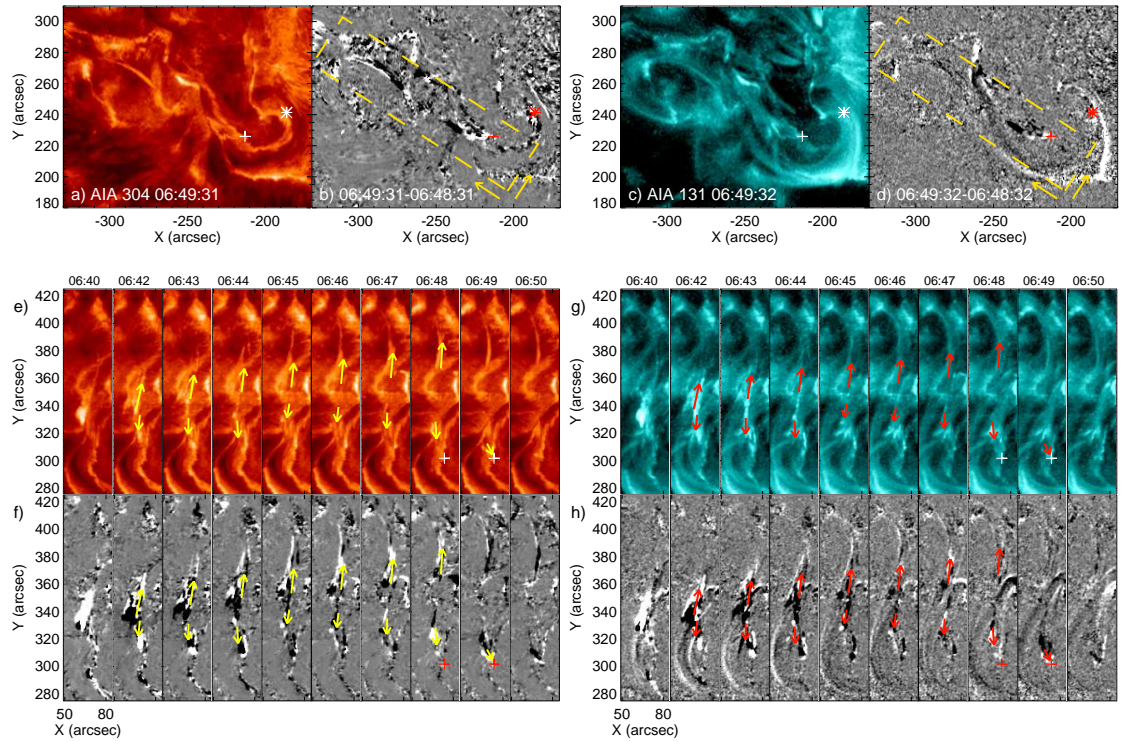}
	\caption{Counter-streaming flows. Panels (a--d) show the sigmoidal loop in AIA 304 and 131~{\AA} original and running difference images, respectively. The `+' symbol marks the termination of the downward flow, while the asterisk marks the final western footpoint of the slipping loop. A rectangular region is cut from both original and running difference images, rotated to the vertical position, and shown in (e--h) in chronological order. A few episodes of counter-streaming flows are marked by arrows. An animation of AIA 304 and 131~{\AA} original and running difference images is available online, covering the time interval of 06:30--07:00 UT. In the animation GOES 1--8~{\AA} lightcurve is provided for context in the top panel, with the time period of slipping motions marked by dashed lines and that of counter-streaming flows by a shaded bar. An animated vertical solid line indicates when the images were taken.
		\label{fig:csflow}}
\end{figure*}

Counter-streaming flows were observed prior to the flare onset, at speeds estimated to be tens of kilometers per second, during a time interval between 06:41--06:49 UT (marked by a shaded bar in Figure \ref{fig:tl}(b \& c)), which coincides with the late phase of slipping motions. In Figure~\ref{fig:csflow} we cut a strip from both original and running difference images, rotate it to the vertical position, and marked a few episodes of flows discernible in the 304 and 131~{\AA} passband by yellow and red arrows, respectively. From the animation accompanying Figure~\ref{fig:csflow}, one can see multiple episodes of paired flows streaming away from where the J-shaped loop apparently crossed the filament. The upward flow is loaded onto, and moves eastward along, the loop, while the downward flow falls back to the surface.
\par
We paid attention to the following observational features: i) the flows consist of filament material that is preferentially visible in AIA 304~{\AA}; ii) both upward and downward flows are released from the similar site and are nearly aligned, but one is directed along the loop and the other along the filament; and iii) the draining site of the downward flow is close to the western footpoint of the original J-shaped loop (plus symbol in Figure~\ref{fig:csflow}) but distinct from that of the full-fledged sigmoidal loop (asterisk symbol in Figure~\ref{fig:csflow}). The last feature, together with the fact that the slipping motions start earlier and last more than two times longer than counter-streaming flows (Figure~\ref{fig:tl}), suggests that the interaction of the loop's eastern segment with the filament is a separate process from the slippage of its western segment. 
\par
These observations indicate that the exchange of mass and magnetic connectivity between the filament and the loop is associated with the release of oppositely directed flows. Thus, we interpreted the counter-streaming flows as the outflows of magnetic reconnection between the filament and the eastern segment of the sigmoidal loop. The reconnection might have facilitated the loop's splitting from the filament. 
\subsection{Evolution of the Source Region}
\begin{figure*}[ht!]
	\centering\includegraphics[width=\textwidth]{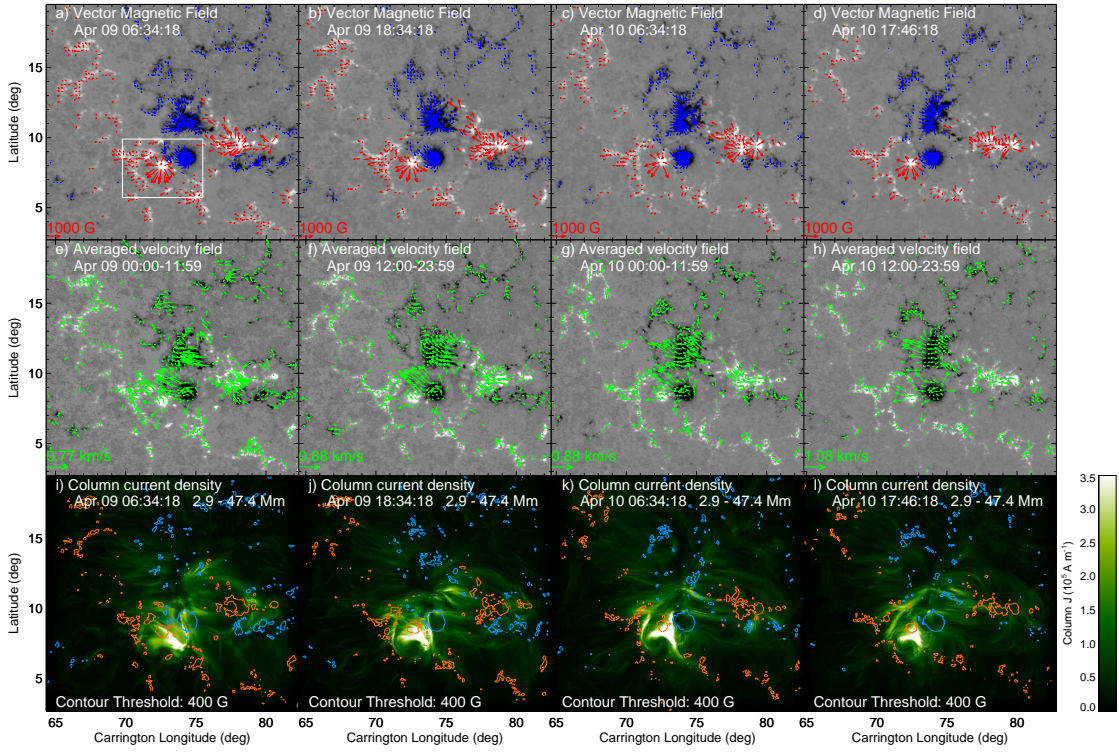}
	\caption{Evolution of the source active region.
		Panels (a)-(d) show vector magnetic field detected by SDO/HMI, with a white rectangle to mark the region from which we obtain the magnetic flux in Figure~\ref{fig:tl}(a).
		In panels (e)-(h), photospheric flow field calculated by DAVE4VM \citep{Schuck2008} is averaged every 12 hours. 
		Panels (i)-(l) show maps of column current density, obtained by integrating the unsigned current density over the same height range as in Figure \ref{fig:mag_cur}(b),  overplotted by orange (blue) contours indicating $B_z$ of $+400$ ($-400$) G.
		\label{fig:fv}}
\end{figure*}

Characterized by flux cancellation (Figures~\ref{fig:tl}(a) and \ref{fig:fv}(a-d)), the evolution of the source region AR 11719, which is categorized as $\beta\gamma$ according to the Mount Wilson classification, is typical of a decayed AR. The magnetic flux in Figure~\ref{fig:tl}(a) is obtained from a rectangle (Figure~\ref{fig:fv}(a)) enclosing the major sunspot pair in the AR, where the slipping motions were mainly observed. 
The flow field calculated by DAVE4VM shows that the two sunspots of opposite polarities both moved eastward, with a small converging motion toward the PIL segment oriented in the southwest-northeast direction (Figure \ref{fig:fv}(e)-(h)). As a result of the photospheric evolution, the column current density became increasingly strengthened and concentrated in the neighborhood of the flaring PIL within one day prior to the flare (Figure \ref{fig:fv}(i--l)). The observed flux cancellation and converging motion in the hosting AR suggests that the photospheric conditions that are favorable for tether-cutting reconnection \cite[]{Ballegooijen1989,Green2009} may also work for the slipping reconnection. \par
\section{Discussion and Conclusion} \label{sec:disc}
To summarize, we investigated the evolution of a J-shaped loop, focusing on the slipping process of its western footpoint and its transformation into a sigmoid. Partially aligned with the middle section of a filament, the J-shaped loop brightened and its western footpoint slipped along the PIL above which the electric current density is enhanced, leading to formation of a hot sigmoidal loop. Meanwhile, this ``single-J to S'' transformation was accompanied by coronal dimming at the fixed footpoint and counter-streaming flows along the loop. \par
\subsection{Slipping Reconnection}
The observation demonstrates that a sequence of slipping reconnection occurred in a low-corona region with enhanced electric current density above the PIL. The electric current density is related to the ``nonidealness'' $\mathbf{R}$ of plasma in terms of Ohm's law, i.e., $\mathbf{R}=\mathbf{E}+\mathbf{v}\times\mathbf{B}$ \citep{Vasyliunas1975}. $\mathbf{R}$ is generally negligible in the large-scale, low-$\beta$ coronal plasma, but can become locally important, where the steep gradients of magnetic field are accompanied by large enhancements in electric current density \cite[]{Vasyliunas1975} as well as in the squashing degree of elemental flux tubes \cite[]{Titov2002,Boozer2012,Janvier2013}. Here we assume that such regions are highly localized and remain unresolved by the telescope, considering that the ion inertial length is on the order of 100~m in the coronal environment. However, any ``microscopic'' slippage of plasma relative to the magnetic field inside these regions would be exponentially amplified \citep{Boozer2012}, due to large squashing factors, into ``macroscopic'' loop motions outside, where the force-free assumption remains valid for the coronal plasma. Therefore, we argue that the observed coronal loops are largely aligned along the coronal magnetic field, despite their visibility depending on plasma temperature and density; hence their motions must represent the change of magnetic configuration. 
\par	
The apparent `footpoints' of coronal loops may be slightly different from where they are truly anchored in the dense chromosphere or photosphere, but they are least affected by the projection effect. Hence, tracking the footpoint motions is useful in diagnosing the reconnection rate in the diffusion region, especially within the framework of the standard 2D flare model \cite[]{Qiu2002}. In 3D scenarios, that field-line footpoints slip at a sub- or super-Alfv\'{e}nic speed is termed slipping or slip-running reconnection \citep{Aulanier2006}. In observational studies, slipping speeds reported in the literature so far are below the acoustic speed in the corona \cite[]{Aulanier2007,Dudik2014,Li2014,Li2015,Dudik2016}. In the present event, the peak slippage speed is as fast as over \kms{1000}, which is comparable with the typical Alfv\'{e}nic speed in the corona. This is the fastest speed ever recorded, probably falling in the slip-running regime.
\par
Once the slippage of plasma relative to the magnetic field is triggered, it would produce an electric field in the reference frame moving with the plasma, which further enhances the nonidealness $\mathbf{R}$. If the increase in $\mathbf{R}$, especially the parallel electric field $E_\parallel$ \citep{Schindler1988}, can help accelerate the slippage, a runaway process would ensue, which might be the case here, as evidenced by the rapid increase in the apparent slipping speed at about 06:40 UT (Figure~\ref{fig:tl}(e)). However, the runaway is not expect to last after the loop slips out of the region of strong current densities, as also suggested by our observation (Figure~\ref{fig:mag_cur}).
\subsection{J-to-S Transformation} \label{discuss:J-to-S}
This event presented several asymmetric features during the transformation from the single J-shaped loop to sigmoid through slipping reconnection. We note that such asymmetric features are not expected nor observed during the transformation from a double J-shaped loop to a sigmoid through tether-cutting reconnection. Coronal dimming was clearly detected around the eastern footpoint, which was relatively stationary, but not quite visible at the western fooptoint undergoing slipping motions, even after the slipping speed had already slowed down. This suggests that the temporal scale for the development of coronal dimming is much longer than that of slipping motions. The asymmetric temperature distribution along the fully fledged S-shaped loop, with its western segment significantly hotter than the eastern one (Figure~\ref{fig:avet}(b), \ref{fig:tl}(f)), resulted most likely from an accumulative effect of the loop segment undergoing slipping motions being heated up on the way. The slipping reconnection obviously proceeded episodically, as evidenced by the bursty profile of the slipping speed (Figure~\ref{fig:tl}(e)), in the $\sim\,$16 min interval during which slipping motions were observed. Hence the temporal scale for slipping reconnection heating must be much smaller than that for plasma cooling, the latter of which is typically on the order of tens of minutes in flare plasmas \citep[e.g.,][their Fig.6]{Chen2020}.
\par
\begin{figure*}[ht!]
	\centering\includegraphics[width=\textwidth]{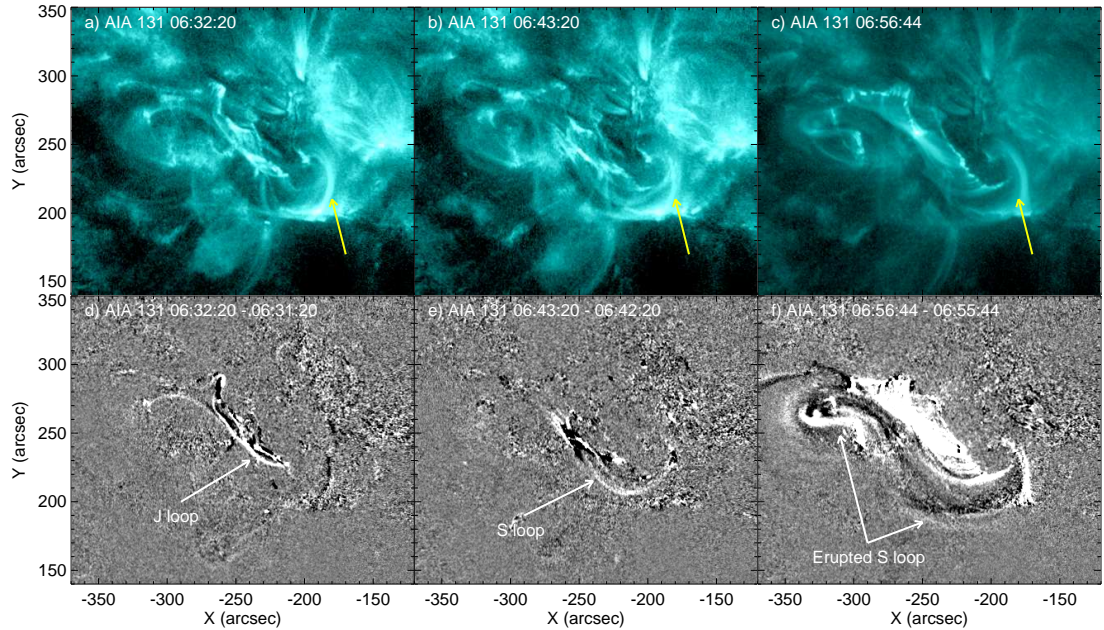}
	\caption{Semicircular loop bundle in retation to the J-to-S transformation. Snapshots of AIA 131~{\AA} original and running difference images are selected to show that the semicircular loop (yellow arrow) remained nearly stationary before slipping motions (left), when the sigmoidal loop was formed (middle), and when the sigmoid erupted (right). 
		\label{fig:semicirc}}
\end{figure*}

The ``single-J to S'' transformation through slipping reconnection distinguishes itself from the ``double-J to S'' transformation through tether-cutting reconnection. Tether-cutting reconnection is so dubbed because it reduces the original four footpoints of the double J-shaped loop to the two footpoints of the S-shaped loop. In a mechanical analogy, the field lines that provide the downward tension are analogous to ground-anchored tethers that hold down a buoyant balloon. In contrast, the slipping reconnection did not `cut' any tethers but continually changed where the J-shaped loop was anchored by shifting one of its footpoints. One might misidentify the semicircular loop in the southwest as the other J-shaped loop in the prototypical picture of the double-J to S transformation. In fact, the semicircular loop was distinct from the loop under slipping motions (Figure~\ref{fig:slip}), although they appear to share the western footpoint by the end of the slipping phase. Further, the semicircular loop remained nearly stationary throughout the J-to-S transformation as well as when the sigmoidal loop started to rise upward to erupt (Figure~\ref{fig:semicirc}), which argues against the involvement of the semicircular loop in the formation of the sigmoid; nor there existed any brightening compact loop straddling the PIL (cf. Figure~\ref{fig:slip}), which argues against the involvement of the standard tether-cutting reconnection in this transient J-to-S transformation. 
\subsection{Structural Complexity}
In addition to the high dynamic complexity, our investigation demonstrates high structural complexity in this active region. It may possess a ``triple decker'' structure, in comparison to the  ``double deckers'' that are more frequently observed \citep[e.g.,][]{Liu2012, Cheng2014double, Awasthi2019, Pan2021}.  Since the blob and the related cusp-shaped loops rose earlier than the S-shaped loop (Figure \ref{fig:tl}(d)) and their temperatures evolved differently from the S-shaped loop, they must be independent of, and lying higher than, the S-shaped loop. On the other hand,  the long-lasting filament remained stationary throughout the slipping process and the subsequent eruption in both H$\alpha$ (Figure \ref{fig:fl}) and 304~{\AA}, suggesting that it is low-lying and relatively independent of the eruption. As a result, the S-shaped loop was `sandwiched' by the underlying filament and the overlying blob. The observation of counter-streaming flows, however, indicates that there existed some interaction between the filament and the sigmoidal loop, most likely through a reconnection event that was isolated from the slipping reconnection but facilitated the splitting of the loop from the filament (\S\ref{subsubsec:counterstreaming}). 
\subsection{Conclusion}
In conclusion, we have demonstrated the pre-flare slipping reconnection of a J-shaped loop system, which slipped through a region of enhanced (column) electric current density in an episodic and bursty fashion, at a speed comparable to the Alfv\'{e}n speed. With one footpoint fixed and the other shifting along the PIL, the loop was transformed into a hot S-shaped loop that served as the core structure for the subsequent eruption. Our analysis therefore suggests a new formation mechanism for ``seed'' flux ropes \citep{Liu2020}, differing from the prototypical scenario with a double-J to S transformation; it also brings new insight into slipping reconnection by revealing the variation of slipping speed, the asymmetric dimming at footpoints, and asymmetric heating along the loop.\par

\acknowledgments
This work was supported by National Natural Science Foundation of China (NSFC 41774150, 11925302, 42188101, and 11903032) and the Strategic Priority Program of the Chinese Academy of Sciences (XDB41030100). SDO is a mission of NASA's Living With a Star Program. H$\alpha$ data were provided by the Kanzelh\"{o}he Solar Observatory, University of Graz, Austria, and by GONG instruments operated by NSO with contribution from NOAA.
\bibliographystyle{aasjournal}
\bibliography{slip}

\begin{thebibliography}{}
\expandafter\ifx\csname natexlab\endcsname\relax\def\natexlab#1{#1}\fi
\providecommand{\url}[1]{\href{#1}{#1}}
\providecommand{\dodoi}[1]{doi:~\href{http://doi.org/#1}{\nolinkurl{#1}}}
\providecommand{\doeprint}[1]{\href{http://ascl.net/#1}{\nolinkurl{http://ascl.net/#1}}}
\providecommand{\doarXiv}[1]{\href{https://arxiv.org/abs/#1}{\nolinkurl{https://arxiv.org/abs/#1}}}

\bibitem[{Aschwanden(2006)}]{Aschwanden2006}
Aschwanden, M. 2006, Physics of the solar corona: an introduction with problems
  and solutions, 2nd edn. (Springer Science \& Business Media)

\bibitem[{{Aulanier} {et~al.}(2006){Aulanier}, {Pariat}, {D{\'e}moulin}, \&
  {DeVore}}]{Aulanier2006}
{Aulanier}, G., {Pariat}, E., {D{\'e}moulin}, P., \& {DeVore}, C.~R. 2006,
  \solphys, 238, 347, \dodoi{10.1007/s11207-006-0230-2}

\bibitem[{{Aulanier} {et~al.}(2007){Aulanier}, {Golub}, {DeLuca}, {Cirtain},
  {Kano}, {Lundquist}, {Narukage}, {Sakao}, \& {Weber}}]{Aulanier2007}
{Aulanier}, G., {Golub}, L., {DeLuca}, E.~E., {et~al.} 2007, Science, 318,
  1588, \dodoi{10.1126/science.1146143}

\bibitem[{{Awasthi} {et~al.}(2019){Awasthi}, {Liu}, \& {Wang}}]{Awasthi2019}
{Awasthi}, A.~K., {Liu}, R., \& {Wang}, Y. 2019, \apj, 872, 109,
  \dodoi{10.3847/1538-4357/aafdad}

\bibitem[{{Boozer}(2012)}]{Boozer2012}
{Boozer}, A.~H. 2012, Physics of Plasmas, 19, 112901, \dodoi{10.1063/1.4765352}

\bibitem[{{Canfield} {et~al.}(1999){Canfield}, {Hudson}, \&
  {McKenzie}}]{Canfield1999}
{Canfield}, R.~C., {Hudson}, H.~S., \& {McKenzie}, D.~E. 1999, \grl, 26, 627,
  \dodoi{10.1029/1999GL900105}

\bibitem[{{Carmichael}(1964)}]{Carmichael1964}
{Carmichael}, H. 1964, {A Process for Flares}, Vol.~50, 451

\bibitem[{{Chen} {et~al.}(2020){Chen}, {Liu}, {Liu}, {Awasthi}, {Zhang},
  {Wang}, \& {Kliem}}]{Chen2020}
{Chen}, J., {Liu}, R., {Liu}, K., {et~al.} 2020, \apj, 890, 158,
  \dodoi{10.3847/1538-4357/ab6def}

\bibitem[{{Cheng} {et~al.}(2014){Cheng}, {Ding}, {Zhang}, {Sun}, {Guo}, {Wang},
  {Kliem}, \& {Deng}}]{Cheng2014double}
{Cheng}, X., {Ding}, M.~D., {Zhang}, J., {et~al.} 2014, \apj, 789, 93,
  \dodoi{10.1088/0004-637X/789/2/93}

\bibitem[{{Cheung} {et~al.}(2015){Cheung}, {Boerner}, {Schrijver}, {Testa},
  {Chen}, {Peter}, \& {Malanushenko}}]{Cheung2015}
{Cheung}, M. C.~M., {Boerner}, P., {Schrijver}, C.~J., {et~al.} 2015, \apj,
  807, 143, \dodoi{10.1088/0004-637X/807/2/143}

\bibitem[{{Cheung} \& {Isobe}(2014)}]{Cheung2014}
{Cheung}, M. C.~M., \& {Isobe}, H. 2014, Living Reviews in Solar Physics, 11,
  3, \dodoi{10.12942/lrsp-2014-3}

\bibitem[{{D{\'e}moulin}(2006)}]{Demoulin2006}
{D{\'e}moulin}, P. 2006, Advances in Space Research, 37, 1269,
  \dodoi{10.1016/j.asr.2005.03.085}

\bibitem[{{Dud{\'\i}k} {et~al.}(2014){Dud{\'\i}k}, {Janvier}, {Aulanier}, {Del
  Zanna}, {Karlick{\'y}}, {Mason}, \& {Schmieder}}]{Dudik2014}
{Dud{\'\i}k}, J., {Janvier}, M., {Aulanier}, G., {et~al.} 2014, \apj, 784, 144,
  \dodoi{10.1088/0004-637X/784/2/144}

\bibitem[{{Dud{\'\i}k} {et~al.}(2016){Dud{\'\i}k}, {Polito}, {Janvier},
  {Mulay}, {Karlick{\'y}}, {Aulanier}, {Del Zanna}, {Dzif{\v{c}}{\'a}kov{\'a}},
  {Mason}, \& {Schmieder}}]{Dudik2016}
{Dud{\'\i}k}, J., {Polito}, V., {Janvier}, M., {et~al.} 2016, \apj, 823, 41,
  \dodoi{10.3847/0004-637X/823/1/41}

\bibitem[{{Fan} \& {Gibson}(2003)}]{Fan2003}
{Fan}, Y., \& {Gibson}, S.~E. 2003, \apjl, 589, L105, \dodoi{10.1086/375834}

\bibitem[{{Gopalswamy} {et~al.}(1999){Gopalswamy}, {Kaiser}, {MacDowall},
  {Reiner}, {Thompson}, \& {Cyr}}]{Gopalswamy1999}
{Gopalswamy}, N., {Kaiser}, M.~L., {MacDowall}, R.~J., {et~al.} 1999, in
  American Institute of Physics Conference Series, Vol. 471, Solar Wind Nine,
  ed. S.~R. {Habbal}, R.~{Esser}, J.~V. {Hollweg}, \& P.~A. {Isenberg},
  641--644

\bibitem[{{Gou} {et~al.}(2016){Gou}, {Liu}, {Wang}, {Liu}, {Zhuang}, {Chen},
  {Zhang}, \& {Liu}}]{Gou2016}
{Gou}, T., {Liu}, R., {Wang}, Y., {et~al.} 2016, \apjl, 821, L28,
  \dodoi{10.3847/2041-8205/821/2/L28}

\bibitem[{{Green} \& {Kliem}(2009)}]{Green2009}
{Green}, L.~M., \& {Kliem}, B. 2009, \apjl, 700, L83,
  \dodoi{10.1088/0004-637X/700/2/L83}

\bibitem[{{Hirayama}(1974)}]{Hirayama1974}
{Hirayama}, T. 1974, \solphys, 34, 323, \dodoi{10.1007/BF00153671}

\bibitem[{{Hoeksema} {et~al.}(2014){Hoeksema}, {Liu}, {Hayashi}, {Sun},
  {Schou}, {Couvidat}, {Norton}, {Bobra}, {Centeno}, {Leka}, {Barnes}, \&
  {Turmon}}]{Hoeksema2014}
{Hoeksema}, J.~T., {Liu}, Y., {Hayashi}, K., {et~al.} 2014, \solphys, 289,
  3483, \dodoi{10.1007/s11207-014-0516-8}

\bibitem[{{Janvier} {et~al.}(2013){Janvier}, {Aulanier}, {Pariat}, \&
  {D{\'e}moulin}}]{Janvier2013}
{Janvier}, M., {Aulanier}, G., {Pariat}, E., \& {D{\'e}moulin}, P. 2013, \aap,
  555, A77, \dodoi{10.1051/0004-6361/201321164}

\bibitem[{{Kopp} \& {Pneuman}(1976)}]{Kopp&Pneuman1976}
{Kopp}, R.~A., \& {Pneuman}, G.~W. 1976, \solphys, 50, 85,
  \dodoi{10.1007/BF00206193}

\bibitem[{{Lemen} {et~al.}(2012){Lemen}, {Title}, {Akin}, {Boerner}, {Chou},
  {Drake}, {Duncan}, {Edwards}, {Friedlaender}, {Heyman}, {Hurlburt}, {Katz},
  {Kushner}, {Levay}, {Lindgren}, {Mathur}, {McFeaters}, {Mitchell}, {Rehse},
  {Schrijver}, {Springer}, {Stern}, {Tarbell}, {Wuelser}, {Wolfson}, {Yanari},
  {Bookbinder}, {Cheimets}, {Caldwell}, {Deluca}, {Gates}, {Golub}, {Park},
  {Podgorski}, {Bush}, {Scherrer}, {Gummin}, {Smith}, {Auker}, {Jerram},
  {Pool}, {Soufli}, {Windt}, {Beardsley}, {Clapp}, {Lang}, \&
  {Waltham}}]{Lemen2012}
{Lemen}, J.~R., {Title}, A.~M., {Akin}, D.~J., {et~al.} 2012, \solphys, 275,
  17, \dodoi{10.1007/s11207-011-9776-8}

\bibitem[{{Li} \& {Zhang}(2014)}]{Li2014}
{Li}, T., \& {Zhang}, J. 2014, \apjl, 791, L13,
  \dodoi{10.1088/2041-8205/791/1/L13}

\bibitem[{{Li} \& {Zhang}(2015)}]{Li2015}
---. 2015, \apjl, 804, L8, \dodoi{10.1088/2041-8205/804/1/L8}

\bibitem[{{Liu} {et~al.}(2013){Liu}, {Deng}, {Lee}, {Wiegelmann}, {Moore}, \&
  {Wang}}]{Liu2013}
{Liu}, C., {Deng}, N., {Lee}, J., {et~al.} 2013, \apjl, 778, L36,
  \dodoi{10.1088/2041-8205/778/2/L36}

\bibitem[{{Liu}(2020)}]{Liu2020}
{Liu}, R. 2020, Research in Astronomy and Astrophysics, 20, 165,
  \dodoi{10.1088/1674-4527/20/10/165}

\bibitem[{{Liu} {et~al.}(2016){Liu}, {Chen}, {Wang}, \& {Liu}}]{Liu2016}
{Liu}, R., {Chen}, J., {Wang}, Y., \& {Liu}, K. 2016, Scientific Reports, 6,
  34021, \dodoi{10.1038/srep34021}

\bibitem[{{Liu} {et~al.}(2012){Liu}, {Kliem}, {T{\"o}r{\"o}k}, {Liu}, {Titov},
  {Lionello}, {Linker}, \& {Wang}}]{Liu2012}
{Liu}, R., {Kliem}, B., {T{\"o}r{\"o}k}, T., {et~al.} 2012, \apj, 756, 59,
  \dodoi{10.1088/0004-637X/756/1/59}

\bibitem[{{Liu} {et~al.}(2010){Liu}, {Liu}, {Wang}, {Deng}, \&
  {Wang}}]{Liu2010}
{Liu}, R., {Liu}, C., {Wang}, S., {Deng}, N., \& {Wang}, H. 2010, \apjl, 725,
  L84, \dodoi{10.1088/2041-8205/725/1/L84}

\bibitem[{{McKenzie} \& {Canfield}(2008)}]{Mckenzie2008}
{McKenzie}, D.~E., \& {Canfield}, R.~C. 2008, \aap, 481, L65,
  \dodoi{10.1051/0004-6361:20079035}

\bibitem[{{Moore} \& {Labonte}(1980)}]{Moore1980}
{Moore}, R.~L., \& {Labonte}, B.~J. 1980, in Solar and Interplanetary Dynamics,
  ed. M.~{Dryer} \& E.~{Tandberg-Hanssen}, Vol.~91, 207--210

\bibitem[{{Moore} {et~al.}(2001){Moore}, {Sterling}, {Hudson}, \&
  {Lemen}}]{Moore2001}
{Moore}, R.~L., {Sterling}, A.~C., {Hudson}, H.~S., \& {Lemen}, J.~R. 2001,
  \apj, 552, 833, \dodoi{10.1086/320559}

\bibitem[{{O'Dwyer} {et~al.}(2010){O'Dwyer}, {Del Zanna}, {Mason}, {Weber}, \&
  {Tripathi}}]{ODwyer2010}
{O'Dwyer}, B., {Del Zanna}, G., {Mason}, H.~E., {Weber}, M.~A., \& {Tripathi},
  D. 2010, \aap, 521, A21, \dodoi{10.1051/0004-6361/201014872}

\bibitem[{{Pan} {et~al.}(2021){Pan}, {Liu}, {Gou}, {Kliem}, {Su}, {Chen}, \&
  {Wang}}]{Pan2021}
{Pan}, H., {Liu}, R., {Gou}, T., {et~al.} 2021, \apj, 909, 32,
  \dodoi{10.3847/1538-4357/abda4e}

\bibitem[{{Patsourakos} {et~al.}(2020){Patsourakos}, {Vourlidas},
  {T{\"o}r{\"o}k}, {Kliem}, {Antiochos}, {Archontis}, {Aulanier}, {Cheng},
  {Chintzoglou}, {Georgoulis}, {Green}, {Leake}, {Moore}, {Nindos}, {Syntelis},
  {Yardley}, {Yurchyshyn}, \& {Zhang}}]{Patsourakos2020}
{Patsourakos}, S., {Vourlidas}, A., {T{\"o}r{\"o}k}, T., {et~al.} 2020, \ssr,
  216, 131, \dodoi{10.1007/s11214-020-00757-9}

\bibitem[{{Pesnell} {et~al.}(2012){Pesnell}, {Thompson}, \&
  {Chamberlin}}]{Pesnell2012}
{Pesnell}, W.~D., {Thompson}, B.~J., \& {Chamberlin}, P.~C. 2012, \solphys,
  275, 3, \dodoi{10.1007/s11207-011-9841-3}

\bibitem[{{Pontin}(2011)}]{Pontin2011}
{Pontin}, D.~I. 2011, Advances in Space Research, 47, 1508,
  \dodoi{10.1016/j.asr.2010.12.022}

\bibitem[{{Priest} \& {D{\'e}moulin}(1995)}]{Priest1995}
{Priest}, E.~R., \& {D{\'e}moulin}, P. 1995, \jgr, 100, 23443,
  \dodoi{10.1029/95JA02740}

\bibitem[{{Priest} {et~al.}(2003){Priest}, {Hornig}, \& {Pontin}}]{Priest2003}
{Priest}, E.~R., {Hornig}, G., \& {Pontin}, D.~I. 2003, Journal of Geophysical
  Research (Space Physics), 108, 1285, \dodoi{10.1029/2002JA009812}

\bibitem[{{Qiu} \& {Cheng}(2017)}]{Qiu2017}
{Qiu}, J., \& {Cheng}, J. 2017, \apjl, 838, L6,
  \dodoi{10.3847/2041-8213/aa6798}

\bibitem[{{Qiu} {et~al.}(2002){Qiu}, {Lee}, {Gary}, \& {Wang}}]{Qiu2002}
{Qiu}, J., {Lee}, J., {Gary}, D.~E., \& {Wang}, H. 2002, \apj, 565, 1335,
  \dodoi{10.1086/324706}

\bibitem[{{Rust} \& {Kumar}(1996)}]{Rust1996}
{Rust}, D.~M., \& {Kumar}, A. 1996, \apjl, 464, L199, \dodoi{10.1086/310118}

\bibitem[{{Sakurai} {et~al.}(1992){Sakurai}, {Shibata}, {Ichimoto}, {Tsuneta},
  \& {Acton}}]{Sakurai1992}
{Sakurai}, T., {Shibata}, K., {Ichimoto}, K., {Tsuneta}, S., \& {Acton}, L.~W.
  1992, \pasj, 44, L123

\bibitem[{{Schindler} {et~al.}(1988){Schindler}, {Hesse}, \&
  {Birn}}]{Schindler1988}
{Schindler}, K., {Hesse}, M., \& {Birn}, J. 1988, \jgr, 93, 5547,
  \dodoi{10.1029/JA093iA06p05547}

\bibitem[{Schuck(2008)}]{Schuck2008}
Schuck, P.~W. 2008, The Astrophysical Journal, 683, 1134,
  \dodoi{10.1086/589434}

\bibitem[{{Sterling} {et~al.}(2000){Sterling}, {Hudson}, {Thompson}, \&
  {Zarro}}]{Sterling2000}
{Sterling}, A.~C., {Hudson}, H.~S., {Thompson}, B.~J., \& {Zarro}, D.~M. 2000,
  \apj, 532, 628, \dodoi{10.1086/308554}

\bibitem[{{Sturrock}(1966)}]{Sturrock1966}
{Sturrock}, P.~A. 1966, \nat, 211, 695, \dodoi{10.1038/211695a0}

\bibitem[{{Su} {et~al.}(2018){Su}, {Veronig}, {Hannah}, {Cheung}, {Dennis},
  {Holman}, {Gan}, \& {Li}}]{Su2018}
{Su}, Y., {Veronig}, A.~M., {Hannah}, I.~G., {et~al.} 2018, \apjl, 856, L17,
  \dodoi{10.3847/2041-8213/aab436}

\bibitem[{{Titov} {et~al.}(2002){Titov}, {Hornig}, \&
  {D{\'e}moulin}}]{Titov2002}
{Titov}, V.~S., {Hornig}, G., \& {D{\'e}moulin}, P. 2002, Journal of
  Geophysical Research (Space Physics), 107, 1164, \dodoi{10.1029/2001JA000278}

\bibitem[{{van Ballegooijen} \& {Martens}(1989)}]{Ballegooijen1989}
{van Ballegooijen}, A.~A., \& {Martens}, P.~C.~H. 1989, \apj, 343, 971,
  \dodoi{10.1086/167766}

\bibitem[{{Vasyliunas}(1975)}]{Vasyliunas1975}
{Vasyliunas}, V.~M. 1975, Reviews of Geophysics and Space Physics, 13, 303,
  \dodoi{10.1029/RG013i001p00303}

\bibitem[{{Vemareddy} \& {Zhang}(2014)}]{Vemareddy2014}
{Vemareddy}, P., \& {Zhang}, J. 2014, \apj, 797, 80,
  \dodoi{10.1088/0004-637X/797/2/80}

\bibitem[{{Wang} {et~al.}(2019){Wang}, {Zhu}, {Qiu}, {Liu}, {Yang}, \&
  {Hu}}]{Wang2019}
{Wang}, W., {Zhu}, C., {Qiu}, J., {et~al.} 2019, \apj, 871, 25,
  \dodoi{10.3847/1538-4357/aaf3ba}

\bibitem[{{Wiegelmann}(2004)}]{Wiegelmann2004}
{Wiegelmann}, T. 2004, \solphys, 219, 87,
  \dodoi{10.1023/B:SOLA.0000021799.39465.36}

\bibitem[{{Wiegelmann} {et~al.}(2012){Wiegelmann}, {Thalmann}, {Inhester},
  {Tadesse}, {Sun}, \& {Hoeksema}}]{Wiegelmann2012}
{Wiegelmann}, T., {Thalmann}, J.~K., {Inhester}, B., {et~al.} 2012, \solphys,
  281, 37, \dodoi{10.1007/s11207-012-9966-z}

\bibitem[{{Zhang} {et~al.}(2017){Zhang}, {Su}, \& {Ji}}]{Zhang2017}
{Zhang}, Q.~M., {Su}, Y.~N., \& {Ji}, H.~S. 2017, \aap, 598, A3,
  \dodoi{10.1051/0004-6361/201629477}

\end{thebibliography}

\end{document}